\documentclass[10pt,conference]{IEEEtran}
\IEEEoverridecommandlockouts
\pdfoutput=1
\usepackage{amsmath,amssymb,amsfonts}
\usepackage{algpseudocode}
\usepackage{algorithm}
\usepackage{graphicx}
\usepackage{textcomp}
\usepackage{braket}
\usepackage{tikz}
\usepackage{hyperref}
\usepackage{subcaption}
\hypersetup{
    colorlinks = true,
    citecolor = magenta,
    linkcolor = purple
}
\begin{document}
\bstctlcite{IEEEexample:BSTcontrol}
\title{Performance of Quantum Networks Using Heterogeneous Link Architectures
\thanks{This work was supported by JST [Moonshot R\&D] [JPMJMS226C].}}

\author{
\IEEEauthorblockN{Kento Samuel Soon\IEEEauthorrefmark{1}\IEEEauthorrefmark{3},
Naphan Benchasattabuse\IEEEauthorrefmark{2}\IEEEauthorrefmark{3},
Michal Hajdu\v{s}ek\IEEEauthorrefmark{2}\IEEEauthorrefmark{3},
Kentaro Teramoto\IEEEauthorrefmark{4},\\
Shota Nagayama\IEEEauthorrefmark{4}\IEEEauthorrefmark{2},
and Rodney Van Meter\IEEEauthorrefmark{1}\IEEEauthorrefmark{3}}\\
\IEEEauthorblockA{\IEEEauthorrefmark{1}\textit{Faculty of Environment and Information Studies, Keio University Shonan Fujisawa Campus, Kanagawa, Japan}}
\IEEEauthorblockA{\IEEEauthorrefmark{2}\textit{Graduate School of Media and Governance, Keio University Shonan Fujisawa Campus, Kanagawa, Japan}}
\IEEEauthorblockA{\IEEEauthorrefmark{3}\textit{Quantum Computing Center, Keio University, Kanagawa, Japan}}
\IEEEauthorblockA{\IEEEauthorrefmark{4}\textit{Mercari R4D, Mercari, Inc., Tokyo, Japan}}
}

\thispagestyle{plain}
\pagestyle{plain}
\maketitle
\begin{abstract}
The heterogeneity of quantum link architectures is an essential theme in designing quantum networks for technological interoperability and possibly performance optimization.
However, the performance of heterogeneously connected quantum links has not yet been addressed.
Here, we investigate the integration of two inherently different technologies, with one link where the photons flow from the nodes toward a device in the middle of the link, and a different link where pairs of photons flow from a device in the middle towards the nodes.
We utilize the quantum internet simulator QuISP to conduct simulations.
We first optimize the existing photon pair protocol for a single link by taking the pulse rate into account.
Here, we find that increasing the pulse rate can actually decrease the overall performance.
Using our optimized links, we demonstrate that heterogeneous networks actually work. Their performance is highly dependent on link configuration, but we observe no significant decrease in generation rate compared to homogeneous networks.
This work provides insights into the phenomena we likely will observe when introducing technological heterogeneity into quantum networks, which is crucial for creating a scalable and robust quantum internetwork.
\end{abstract}


\section{Introduction}
Distributing Bell pairs between any arbitrary locations we want is a crucial issue in quantum information systems~\cite{wehner2018quantum,rfc9340,azuma2023quantum,hajdusek2023quantum}.
Many critical real-world applications exist, such as information-theoretically-secure quantum cryptography~\cite{pirandola2020advances}, precise quantum sensing~\cite{proctor2018multiparameter,gottesman2012longer}, distributed and blind quantum computation~\cite{caleffi2022distributed, fitzsimons2017private}, and high-speed distributed consensus algorithms~\cite{ben2005fast}.
There also exist experimental feats that demonstrate long-distance entanglement distribution—see Young \emph{et al.} for a recent summary of those experiments~\cite{young2022architecture}.

However, generating entanglement over a long distance or even in a complex, large-scale data center network is difficult due to inherent fiber attenuation.
In classical communication, it is conventional to copy and resend the data in the middle of a link with the help of repeaters, but for the quantum case, we cannot use the same approach due to the no-cloning theorem~\cite{wootters1982cloning,park1970concept}.
One promising method for entanglement distribution is generating link-level Bell pairs and utilizing quantum repeaters to perform entanglement swapping~\cite{zukowski1993event} and entanglement purification~\cite{briegel1998quantum}, expanding a set of link-level entanglement shared between two adjacent nodes into a long-length distributed entanglement.
Utilizing such repeaters allows for dealing with photon loss and performing error management to distribute high-quality Bell pairs.

In order to efficiently generate link-level Bell pairs, a systematic architecture of link-level Bell pairs, namely, quantum link architectures, needs to be addressed.
Jones \emph{et al.} conducted a high-level theoretical analysis among memory-based link architectures~\cite{jones2016design}. They discovered that the choice of link architectures can significantly affect the generation rate, depending on the hardware parameters.
Additionally, Soon \emph{et al.} extends the link architecture utilizing the \emph{entangled photon pair source} (EPPS) by offering practical implementation details and demonstrates that the generation time saturates after reaching a certain quantum memory capacity~\cite{soon2024msm}.
Azuma \emph{et al.} proposed all-photonic repeaters~\cite{azuma2015all}, which are quantum link architectures that do not specifically use memory qubits.
All-photonic architectures introduce redundancy in the transmitted photonic states, which leads to near-deterministic entanglement generation.

A quantum network does not need to limit itself to a single link architecture.
A proper combination of link architectures can play a crucial role in enhancing the performance of a data center-sized quantum multicomputer or crossing the boundaries from a multicomputer's internal network to an external one.
Moreover, as seen in the evolution of the Internet, it is hard to control the independent deployment of technologies, and we are likely to have the same issues in building and scaling up a quantum internet. 
It can fairly be assumed that various organizations will utilize different technologies, and we need to deal with combining them one day.
Work has been done on heterogeneous networks for quantum key distribution (QKD) networks, where they use satellites and field fiber for constructing such links~\cite{chen2021integrated}.

Therefore, addressing the heterogeneity of quantum link architectures is essential.
Existing work on quantum link architectures and quantum repeater networks has studied homogeneous paths~\cite{jones2016design} and heterogeneous paths~\cite{van2013path}.
However, the work on heterogeneous paths did not investigate the consequences of introducing heterogeneity into the link architectures used in a quantum internet, especially involving link architectures utilizing different combinations of optical components.

In this research, we evaluate the performance of the heterogeneous networks consisting of Memory-Interference-Memory (MIM) and Memory-Source-Memory (MSM) links, which are quantum link architectures that have difference in the combinations of optical components.
In the former link architecture, photons flow from the nodes toward a device in the middle of the link, and in the latter architecture, photons flow from a device in the middle towards the nodes.
Here, heterogeneity regarding all-photonic repeaters and other quantum links that do not specifically use quantum memory is outside our scope and left for future work.
We conduct simulations utilizing the quantum internet simulator QuISP~\cite{satoh2022quisp}.
In order to test the MSM link fairly against the MIM link, we extend the analysis on the MSM link to consider the entangled photon pair pulse rate.
Our simulations show that increasing the pulse rate does not necessarily increase the Bell pair generation rate and can even decrease it.
Furthermore, we perform simulations to show that heterogeneous networks can successfully implement entanglement swapping.
Performance is highly dependent on link configuration, but the generation rate does not significantly change compared to homogeneous links.

The insights we have gained from these simulations — such as excessive EPPS pulse rate resulting in less performance in MSM links, and worst-link dependency of the performance of heterogeneous links, are the behaviors we likely will observe in real quantum networks as technology evolves.  The ability to thrive in diverse cases is a hallmark of a robust architecture, and we believe it to be crucial for developing a scalable and robust quantum internetwork.
\section{Preliminaries}
This section presents an overview of the essential background required to comprehend this paper and the specifications for our quantum internet simulator, QuISP.

\subsection{Quantum networks}
In both quantum and classical communication, we utilize light to send signals. However, light propagating through optical fiber is subject to attenuation, even in ideal conditions. Therefore, in classical communication, it is conventional to copy, amplify, and resend the data using a repeater to reach the desired location without losing any information. On the other hand, such an approach is impossible when dealing with quantum information due to the no-cloning theorem~\cite{wootters1982cloning}.

To overcome this issue, quantum repeaters~\cite{briegel1998quantum} divide a long end-to-end channel that is unlikely to distribute a Bell pair into reasonably short links and apply entanglement swapping~\cite{zukowski1993event} to establish an end-to-end Bell pair.
Consider an example where we have three quantum network nodes, which we label as $A$, $B$, and $C$.
Here, $A$, $B$, and $C$ are nodes equipped with quantum memories.
We share a link-level entanglement in the quantum state $\ket{\Phi^+}_{AB_1}=(\ket{00}+\ket{11})/\sqrt{2}$ between $A$ and $B$, and $\ket{\Phi^+}_{B_2C}=(\ket{00}+\ket{11})/\sqrt{2}$ between $B$ and $C$.
Under this condition, $B$ measures the two qubits it owns on the Bell state basis. Then, $B$ sends the measurement result via a classical channel.
Once $A$ and $C$ receive this message, they apply corresponding Pauli operations to collapse the quantum state into a predetermined entangled state.
This results in sharing a Bell pair among $A$ and $C$.

\subsection{Quantum link architectures}
Above, we described entanglement swapping using memory-based operations, which is a deterministic operation.
In many link architectures, we utilize optical entanglement swapping, via the inherently probabilistic photonic Bell state measurement (BSM).
This suggests that if linear optical components are used in this link-level entanglement generation, even trying to exclude every possible noise/loss source, there is an ideal probability of failing at $50\%$~\cite{calsamiglia2001optical}.
\subsubsection{Memory-Interference-Memory (MIM) link}
\begin{figure}[htbp]
    \centering
    \includegraphics[keepaspectratio, width=0.45\textwidth]{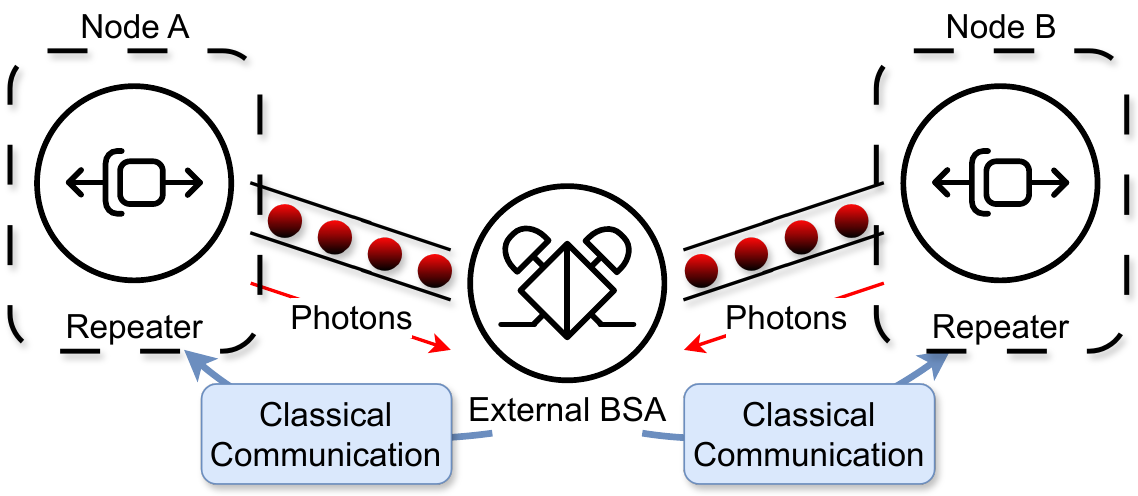}
    \caption{
    MIM link architecture. An external BSA is located between the two nodes. The two nodes emit photons from all available memories. The memories are then locked up until classical message response from the BSA is received. The classical message contains a list of BSM result, indicating the of success or failure, and the Pauli frame correction operations to apply to the memories upon success.
}
    \label{fig:mim}
\end{figure}

One way to create distant node entanglement is by using a Bell state analyzer (BSA), placing it between two nodes, and sending back the results of the BSM to the local nodes, as shown in Fig.~\ref{fig:mim}. With these results, the nodes determine whether to discard the corresponding qubit and, if they keep it, whether to apply the quantum operation to correct the state so that we result in one specific Bell pair, $\ket{\Phi^+}$ state. This work was originally proposed in \cite{simon_robust_2003,duan_efficient_2003,feng_entangling_2003}.
\subsubsection{Memory-Memory (MM) link}
\begin{figure}[htbp]
    \centering
    \includegraphics[keepaspectratio, width=0.45\textwidth]{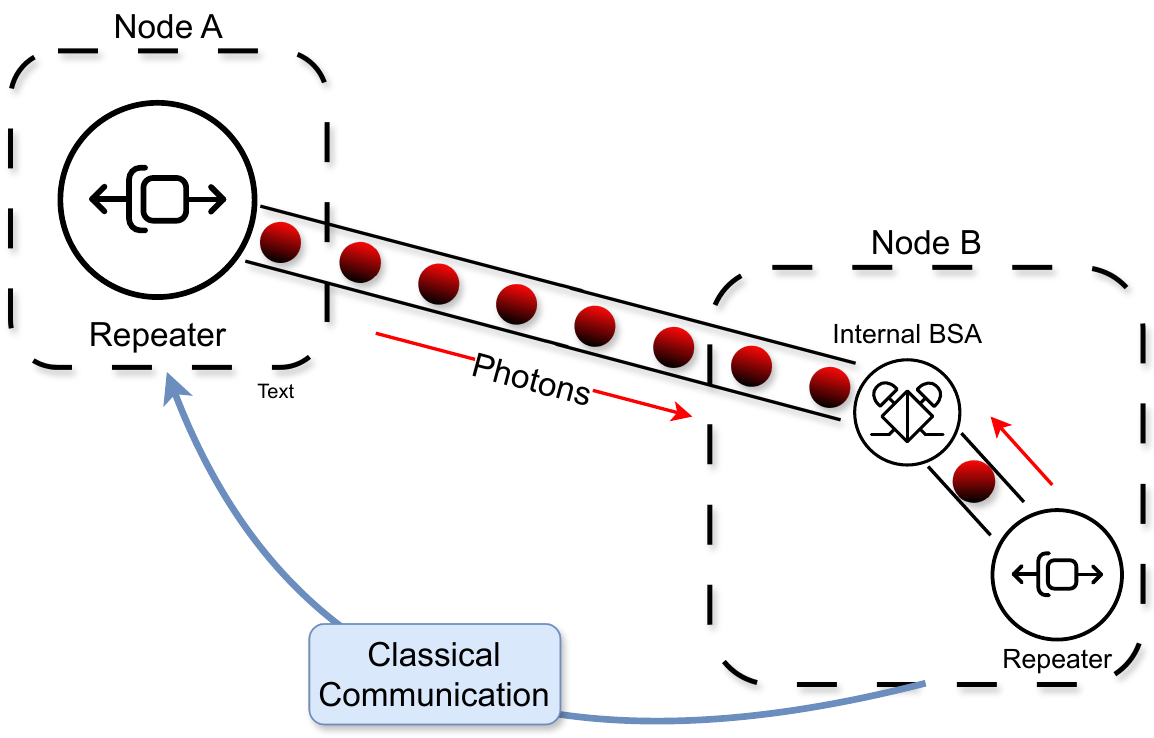}
    \caption{MM link architecture with an internal BSA at one of the nodes. The basic protocol is generally the same for MIM links, but since the node on one side is equipped with the internal BSA, it can immediately decide whether to reset or keep a memory locked.}
    \label{fig:mm}
\end{figure}

In the Memory-Memory link architecture~\cite{munro_quantum_2010}, the essential components are more or less the same as the MIM link, but the BSA is moved inside one of the nodes, as shown in Fig.~\ref{fig:mm}.
The side equipped with a BSA can make immediate decisions on whether to keep a memory locked or to free it since it can quickly assess the success or failure of a BSM.
However, in MM, where we interpret it as a link architecture sending photons from one side to another, optical BSA is not the only possible technology as one might utilize absorptive memories~\cite{lago2021telecom} or other various technologies~\cite{van2008hybrid}.

\subsubsection{Memory-Source-Memory (MSM) link}
\begin{figure*}[h]
    \centering
    \includegraphics[keepaspectratio, width=\textwidth]{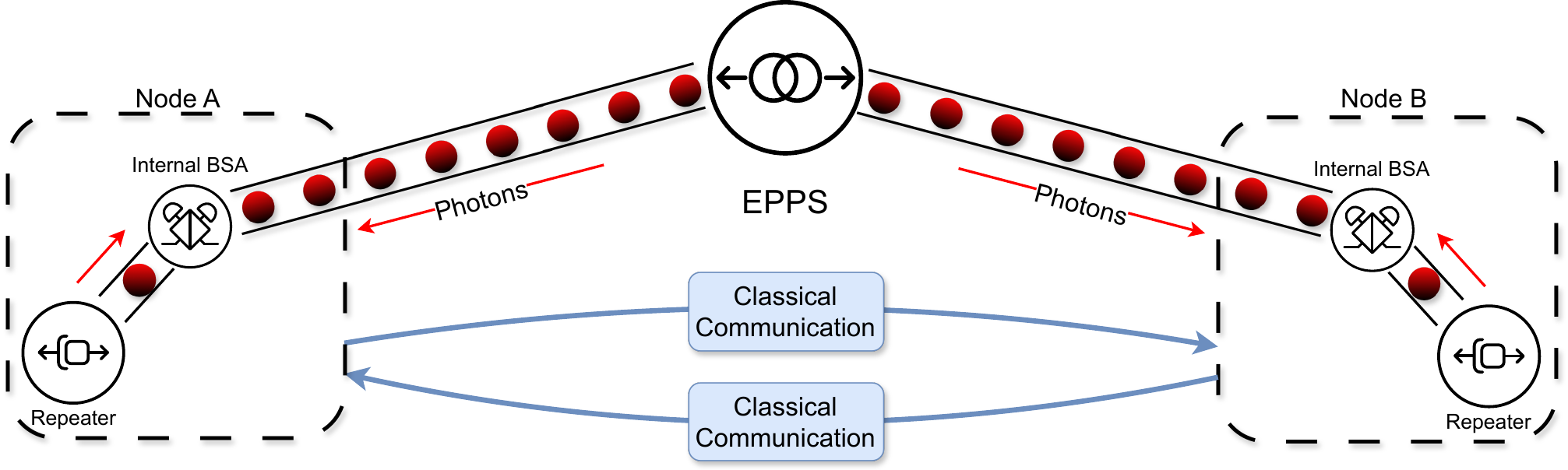}
    \caption{MSM link architecture. An external EPPS (Entangled Photon Pair Source) is between the two nodes. The nodes actively make decisions regarding whether to retain or discard memories and what post-processing to apply in response to the measurement result or classical message they receive. The EPPS continues emitting photons at a fixed pulse rate, and each node independently counts incoming photon pairs, allowing individual partners to identify which photon pair their qubits are entangled with. If we successfully perform a local BSM, we retain the memory associated with the entangled photon and notify our partner node of the success on that photon pair, along with the corresponding Pauli frame. If we fail in a local BSM, we reset the memory and attempt again, sending a failure message to our partner. Additionally, when the memory is fully occupied, we disregard further incoming photons and notify our partner that the BSM has also failed for those photon pairs.}
    \label{fig:msm}
\end{figure*}

Another way to create distant node entanglement is by inserting an entangled photon pair source (EPPS), as shown in Fig.~\ref{fig:msm}. This was first introduced in \cite{jones2013high}. In our definition, the MSM link architecture is partially similar to the MM link architecture, as the nodes can make efficient decisions and allow such nodes to be on both sides using an EPPS, but the overall success rate is lower since two optical BSAs are used, rather than just one.

Previous research have laid out concrete protocols to fully utilize the theoretical predictions made in~\cite{jones2016design} regarding MSM link architecture~\cite{soon2024msm}.
Using this protocol, they perform simulations on a single link and compares its behavior against the conventional MIM protocol.
They observe that the MSM protocol performs much more efficiently in cases where there are sufficient amounts of quantum memories.

However, a saturation effect was also observed: even if the number of memories in our node increases, it inevitably reaches a limit at which the entanglement generation rate does not improve.
Let $L$ denote the distance between the EPPS and the node, $c_\text{fiber}$ represent the speed of light in the fiber, $p_\text{success}$ indicate the probability of a local node succeeding in BSM, and $f_\text{EPPS}$ represent the EPPS pulse rate.
By denoting the number of quantum memories as $\mathcal{N}$, the saturation effect is observed as not providing any speedup in Bell pair generation time where
\begin{equation}
\mathcal{N} \ge \left\lceil\frac{2L}{c_\text{fiber}}p_\text{success}f_\text{EPPS}\right\rceil.
\label{saturation}
\end{equation}

\section{Tuning MSM link performance}
Previously, Soon \emph{et al.} considered the MSM link architecture to have a fixed EPPS pulse rate~\cite{soon2024msm}.
However, the EPPS pulse rate in MSM links was not discussed in detail and was arbitrarily fixed to match the BSA detection rate.
Here, we will adjust the pulse rate to be close to the optimal value with the following procedures.

First, the EPPS collects the number of memories and the success probability of Bell state measurements for each node. 
From these values, we determine the optimal rate of entangled photon pair emissions, taking into account \eqref{saturation}.
However, we also need to consider the BSA detection repetition limit.
This depends on the detector recovery time and classical electronics in the BSA.
The optimal EPPS pulse rate can then be derived as 
\begin{equation}
    f_\text{EPPS} = \min\left(\left\lceil\frac{\mathcal{N}_\text{left} c_{\text{fiber}}}{2 p_\text{left} L_\text{left}}\right\rceil,\left\lceil\frac{\mathcal{N}_\text{right} c_{\text{fiber}}}{2 p_\text{right} L_\text{right}}\right\rceil,f_\text{BSA}\right)\label{adaptiveepps},
\end{equation}
where $f_\text{BSA}$ is the maximum detection rate of the BSA, $\mathcal{N}$ is the number of memories, $p$ is the success probability, and $L$ is the distance from the node to the EPPS with subscripts denoting the position of the node with respect to the EPPS.
This optimal rate also accounts for cases where the EPPS is situated in an imbalanced location.
Our simulation results will show the differences between the adaptive and non-adaptive versions of MSM links.

\section{Heterogeneous Networks}

In classical networks, the layered protocol model has enabled independent development of each layer~\cite{cerf1974protocol, clark2018design}.
Thanks to the abstraction provided by such architecture, heterogeneity in utilizing inherently different technologies naturally arose, and most network engineers above the physical layer can ignore this difference when operating such networks and interconnected networks.
In the quantum case, there are many different levels of heterogeneity, both at the hardware level, where various technologies are used to implement qubits and quantum link architectures and at the software level, where different error correction techniques are utilized.
We focus on hardware-level heterogeneity between link architectures with different underlying technologies, namely, the MIM and MSM links.

Although these two link types are considered significant components of establishing a quantum network, the behavior and performance of what happens when we mix MIM and MSM links, or the heterogeneity of these two links, is still not addressed.
Therefore, this work analyzes end-to-end Bell pair generation performance of a network composed of quantum networks.
This aims to create a genuinely abstracted, robust quantum network of quantum networks, a quantum internet.

We now consider the two major differences between these two types of links.
First, the generation time differs.
This is natural because different mechanisms of entanglement generation are employed.
Second, the way stationary qubits are consumed in the memory is also different.
In MIM, once the entire memory of the node has finished emitting the photons, the BSA sends back an array of BSM results, and then the local nodes apply the corresponding correction operations.
Similar procedure is done for MSM, but the BSM results are sent back individually for each photon pair instead of a batch.

Though there might be some other interesting networks that can be discussed in the sense of heterogeneity, to narrow down our scope, we limit them to the following two networks; (1) a simple entanglement swapping network consisting of two-hop nodes where we freely swap the link architectures, and (2) a network chain consisting mainly of MSM links but with a single MIM link inserted.
\section{Methods}
In order to determine the performance of heterogeneous networks, we utilize the quantum internet simulator QuISP~\cite{satoh2022quisp}.

\subsection{QuISP specific clarifications}
\begin{figure*}[h]
    \centering
    \includegraphics[keepaspectratio, width=\textwidth]{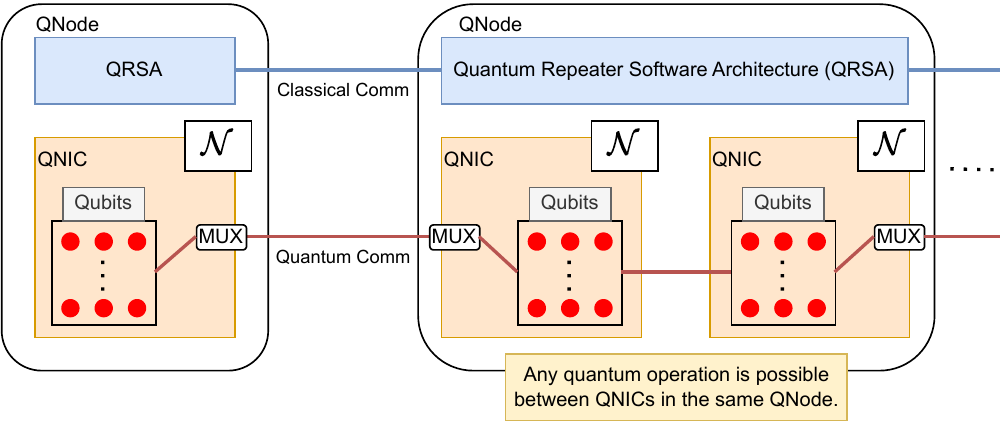}
    \caption{The QuISP architecture of QNICs, QNodes (Quantum Networking Nodes), and quantum memories. Quantum memories can be set per QNIC. Here, the number of qubits is set to $\mathcal{N}$, and we can perform any single or multiple qubit gates within the quantum memories in the same node.}
    \label{fig:nodeQNICarchitecture}
\end{figure*}

In our simulation model, the quantum repeater nodes have a Quantum Network Interface Card (QNIC) with $\mathcal{N}$ memories for each adjacent link.
Therefore, the node acting as a repeater has a total of $2\mathcal{N}$ memories, where CNOT gates can be applied between any pair of qubits even if they reside in different QNIC, as shown in Fig.~\ref{fig:nodeQNICarchitecture}.

In QuISP and in the testbed network we are constructing, the order of entanglement swapping depends on what we specify in the RuleSet~\cite{satoh2022quisp,vanmeter2022a,teramoto2023ruleset, satoh2023rula}, which provides a comprehensive definition of a set of commands that instruct what each node should do, depending \emph{only} on local events and message arrival.
Many approaches to sequencing entanglement swapping are possible. The default entanglement swapping scheme used in QuISP is by a bisection method, where the end-to-end path is recursively split in the middle to maximize the parallelization of swaps at every time step.
This swapping method takes $\mathcal{O}(\log{N})$ timesteps, where $N$ is the number of nodes.

\subsection{Default configurations for our simulations}
Here, we lay out the default configurations for the parameters our links have in common across all simulations.
The major parameters are set as follows.

\begin{itemize}
    \item Attenuation rate in the optical fiber is $0.2$ dB/km.
    \item Speed of light in optical fiber is defined by $c_\text{fiber} = 208189$km/s.
    \item BSA success probability is defined by $p_\text{BSA}=0.5$ (optimal value).
    \item EPPS pulse rate is defined by $f_\text{EPPS}=1$MHz (for non-adaptive MSM protocols).
    \item BSA detection rate is defined by $f_\text{BSA}=1$MHz.
\end{itemize}
The EPPS pulse rate and the BSA detection rate were selected in order to keep the execution time of the simulations feasible.
The support nodes (EPPS or BSA) are placed right in the middle between adjacent nodes.

In our simulation, we measure the time to generate 100 Bell pairs, utilizing the Bell pair generation rate (BP/s) as our metric. 
These Bell pairs are generated between the network's two end nodes. However, discrepancies arise in the recorded time required to generate 100 Bell pairs due to local information at each end node.
Each end node marks the total Bell pair generation time upon receiving the corresponding classical messages.
However, in reality, a Bell pair is genuinely shared between the remote nodes only when both have received their respective classical messages. Therefore, we record the later timing between the two nodes for Bell pair generation time.

We performed simulations 100 times for each simulation instance, and the error bars of the figures represent the standard deviation of the data, barely visible in most cases.

\subsection{Outline of our simulations}
We focused on the following situations, in order to test the adaptive frequency of the MSM link and the heterogeneity introduced into the quantum network.

First, we conducted an experiment comparing the performance of adaptive MSM links, non-adaptive MSM links, and conventional MIM links, varying the number of memories in each QNIC (\textbf{Experiment 1}).

\begin{figure}[htbp]
    \centering
    \includegraphics[keepaspectratio, width=0.45\textwidth]{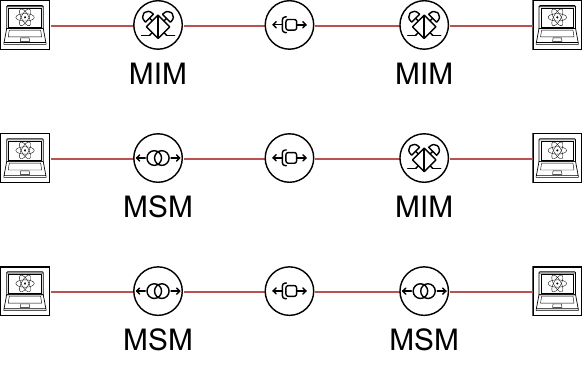}
    \caption{Two hop networks where the link architectures are of various mixtures, which we simulated in experiment two.}
    \label{fig:mimmim_msmmim_msmmsm}
\end{figure}

Second, we conducted another experiment comparing the performance of a two-hop network with homogeneous and heterogeneous links, as illustrated in Fig.~\ref{fig:mimmim_msmmim_msmmsm}, varying the number of memories in each QNIC (\textbf{Experiment 2}).

\begin{figure*}[htbp]
    \centering
    \includegraphics[keepaspectratio, width=\textwidth]{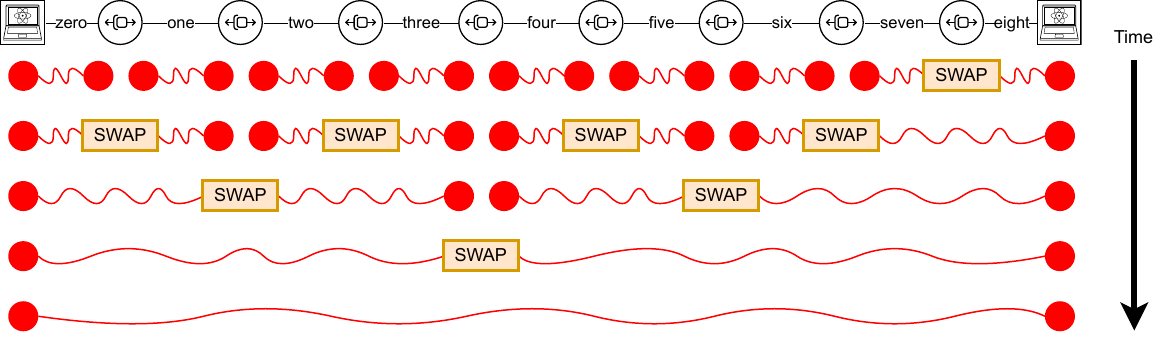}
    \caption{The 10 node chain network. We name the links as shown in the top line, labeling them from link zero to link ten. They are all MSM links, but an MIM link replaces one specified link except for the homogeneous case. The lower portion of the figure illustrates the entanglement swapping sequence used in our simulations.}
    \label{fig:chain_ntwk}
\end{figure*}

Finally, we conducted an experiment to investigate whether replacing segments of a homogeneous path with different link architectures, thereby creating a heterogeneous path as depicted in Fig.~\ref{fig:chain_ntwk}, affects performance (\textbf{Experiment 3}).


While our simulations focus on these three instances, they lay the groundwork for adaptive MSM links and heterogeneous networks.
This framework provides a foundation for future investigations, such as integrating multiplexing of links into the networks or exploring heterogeneity among an even wider array of link architectures, to build upon our findings.

\section{Results}
In this section, we present our observations and explanations for the simulation cases we have listed above.  

\subsection*{\textbf{Experiment 1: Optimizing single MSM link performance}}

\begin{figure*}
\begin{minipage}[h]{0.5\textwidth}
    \centering
    \includegraphics[keepaspectratio, width=\textwidth]{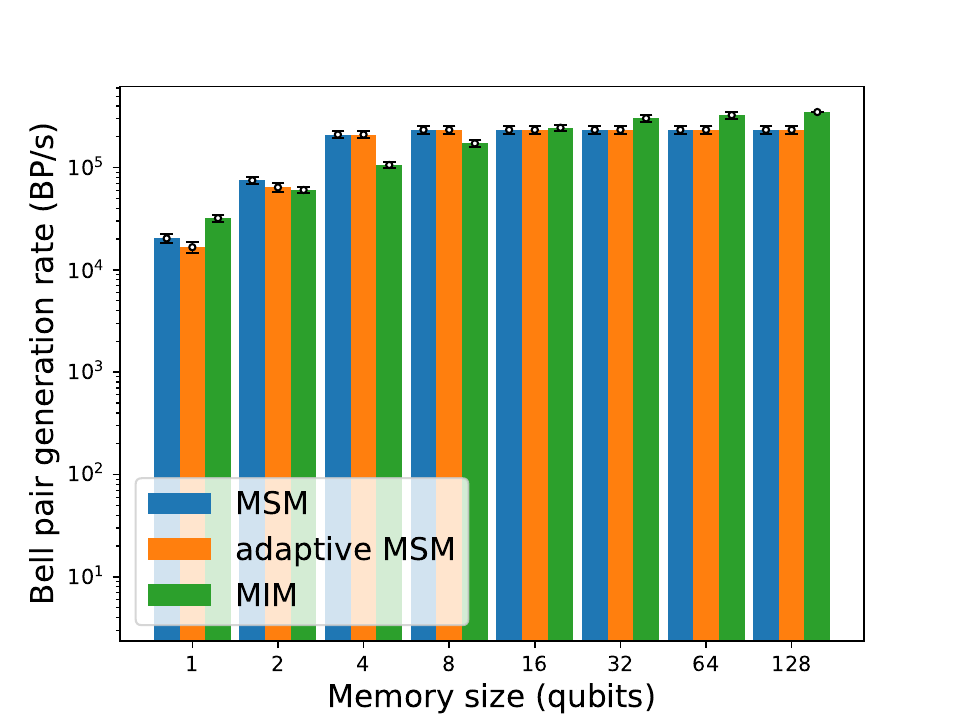}
    \subcaption{1km distance between nodes.}
    \label{fig:bellpair1}
\end{minipage}
\begin{minipage}[h]{0.5\textwidth}
    \centering
    \includegraphics[keepaspectratio, width=\textwidth]{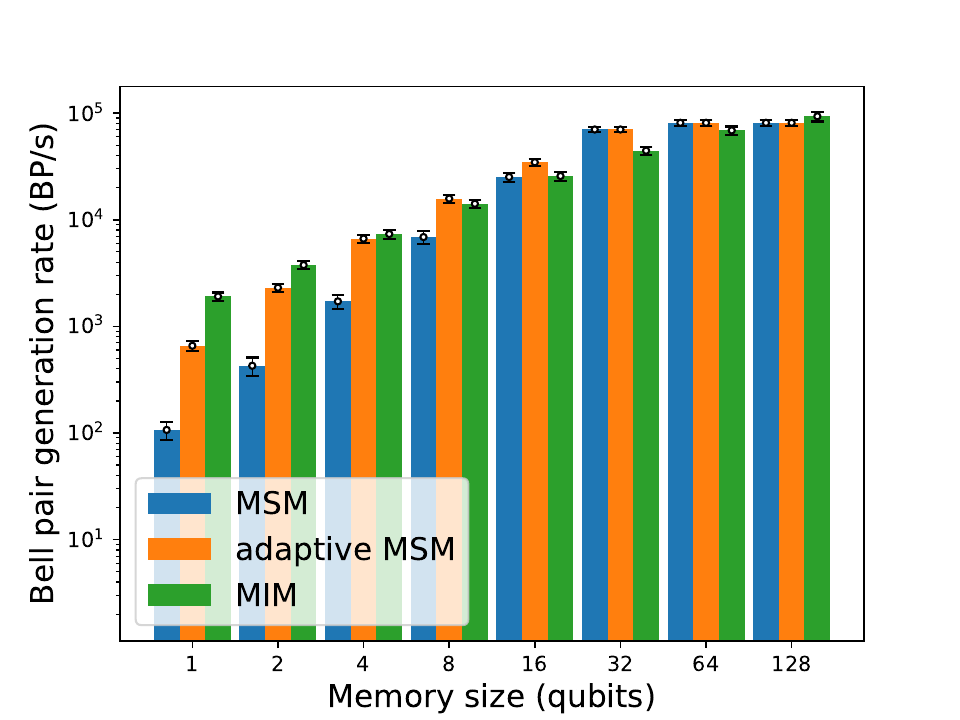}
    \subcaption{20km distance between nodes}
    \label{fig:bellpair20}
\end{minipage}
\caption{Experiment 1: Number of qubits in quantum memory vs Bell pair generation rate for single MSM links, single Adaptive MSM links, and single MIM links, over 1km or 20km. The vertical axis is logarithmic, so the performance differences are large. Notice that in Fig.~\ref{fig:bellpair1}, the overall generation rate between adaptive and non-adaptive MSM links is small, and in Fig.~\ref{fig:bellpair20}, the overall generation rate between adaptive and non-adaptive MSM links is in the order of magnitude in low memory regions, but as soon as the adaptive EPPS rate adjustment gets limited by the BSA rate, the difference between non-adaptive MSM disappears. Also, take notice that the MIM link increases its performance linearly with the number of quantum memories, but for either MSM or adaptive MSM, that is not the case.}
\label{fig:experiment1}
\end{figure*}

First, we conducted a simulation to investigate the effects resulting from alterations in the EPPS pulse rate, as outlined in equation \eqref{adaptiveepps}. 
We adjusted the number of memories within the QNICs, keeping the same node separation distances set at either 1km or 20km. The results are shown in Figs.~\ref{fig:bellpair1} (1km) and~\ref{fig:bellpair20} (20km).

\subsubsection{MIM vs. MSM, Adaptive MSM}
We can observe that for MIM links, the performance increases linearly with the number of quantum memories.
This relationship stems from the direct dependence of MIM link performance on the number of photons emitted in each trial, which in turn is constrained solely by the number of memories available.
In contrast, the performance improvement in MSM links (both adaptive and non-adaptive) is not linear. 
Unlike MIM links, MSM links are constrained by the EPPS pulse rates governing the number of photons per trial, with memory count at end nodes representing another capacity limit for Bell pair distribution if the number of memories is lower. Furthermore, we observe a saturation effect in both adaptive and non-adaptive protocols, where the generation rate plateaus beyond a certain threshold, denoted as $\mathcal{N}$. Even in the adaptive scenario, where we regulate the EPPS pulse rate to prevent surpassing the BSA detection rate, this saturation phenomenon persists.

\subsubsection{MSM vs Adaptive MSM}

When the node-to-node distance is set to 1km, opting for the adaptive pulse rate leads to a decrease in the Bell pair generation rate. 
Notably, the saturation effect becomes evident with the increase in the number of quantum memories.
However, the distinction between adaptive and non-adaptive MSM links is statistically insignificant, thus employing the adaptive protocol on short links appears to have minimal impact.

However, when extending the node-to-node distance to 20km, the adaptive protocol demonstrates a significant improvement, achieving an order of magnitude improvement over the non-adaptive counterpart. 
To calculate the predicted optimal adaptive EPPS pulse rate ($f_{\text{EPPS}}$) for a memory size of $\mathcal{N}=1$, we use the formula $p_\text{success}=p_\text{BSA} p_\text{fiber}$, where $p_\text{fiber}=e^{-L/L_0}$ with attenuation length $L_0=21$km and $L=10$km, yielding $p_\text{success}=0.3106$ and $f_{\text{EPPS}}\simeq 33517\text{Hz}$. 

\begin{figure}[htbp]
    \centering
    \includegraphics[keepaspectratio, width=0.4\textwidth]{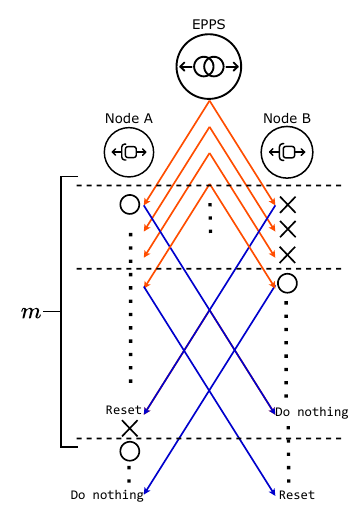}
    \caption{Demonstration of the mutual latch-fail phenomena of incoming entangled photon pairs. The red lines represent entangled photon pairs, and the blue lines represent classical communication. For clarity, we omitted the trivial messages in this diagram.}
    \label{fig:miscomm}
\end{figure}

In contrast, for a non-adaptive MSM link, $f_\text{EPPS}$ was set to $1\text{MHz}$. 
Surprisingly, despite reducing the EPPS pulse rate by a factor of 30, we observed an enhancement in overall entanglement generation. This counterintuitive phenomenon is further demonstrated in Fig.~\ref{fig:miscomm}.

Here, the red line represents entangled photon pairs, the blue lines represent classical communication, and the circles and crosses represent the local BSM success and failure.
Let the two nodes of interest be node $A$ and node $B$, and consider the case where $A$ has succeeded and $B$ failed.
In this case, the photon pair that $A$ has latched is already invalid, but we do not know that until we receive the classical message from $B$.
While waiting for that classical message, there will be $m$ events of photon emission trials from the EPPS,
\begin{equation}
    m = \left\lfloor f_\text{EPPS}\frac{2L}{c_\text{fiber}}\right\rfloor.
\end{equation}
With this $m$, the probability that we observe at least one success in $B$ after its failure on the photon pair which succeeded on $A$ is denoted by $p_{\geq 1}$, where
\begin{equation}
    p_{\geq 1} = 1-(1-p_\text{success})^m.
\end{equation}
However, the photon pair that $B$ has latched is invalid, as observed in the initial case.
It is natural to think that this phenomenon occurs repeatedly.
Hence, when the probability $p_{\geq 1}$ is excessively high, we risk encountering a recurring obstacle where this phenomenon continues. Thus, establishing link-level entanglement could become time-consuming, particularly when $p_{\geq 1}$ approaches one.
Therefore, ensuring that this probability remains sufficiently far from one is crucial for facilitating efficient entanglement generation in low-memory count MSM links.

To validate our analysis, we substitute the parameters for the 20km link to see the values of these probabilities.
Take note that in this case (where the node distance is set to be 20km), the success probability for each BSM is obtained as $p_\text{success}=0.31$.
Calculating the probability $p_{\geq 1}$ for $f_\text{EPPS}=1\text{MHz}$ and $m=96$, we obtain asymptotically $p_{\geq 1}=1$, and for $f_\text{EPPS}=33517\text{Hz}$ and $m=3$, we obtain $p_{\geq 1}\simeq 0.67$.
This analysis suggests that having excessive EPPS pulse rate can decrease the entanglement generation rate, thereby providing a clear explanation for the observed behavior.

\subsection*{\textbf{Experiment 2: Two Hops}}
We now analyze two-hop networks where entanglement swapping comes into action by comparing networks that include both MSM and MIM links (heterogeneous) against those that consist solely of MSM or MIM links (homogeneous), as shown in Fig.~\ref{fig:mimmim_msmmim_msmmsm}.

\begin{figure}[h]
\begin{minipage}[h]{0.5\textwidth}
    \centering
\includegraphics[keepaspectratio, width=\textwidth]{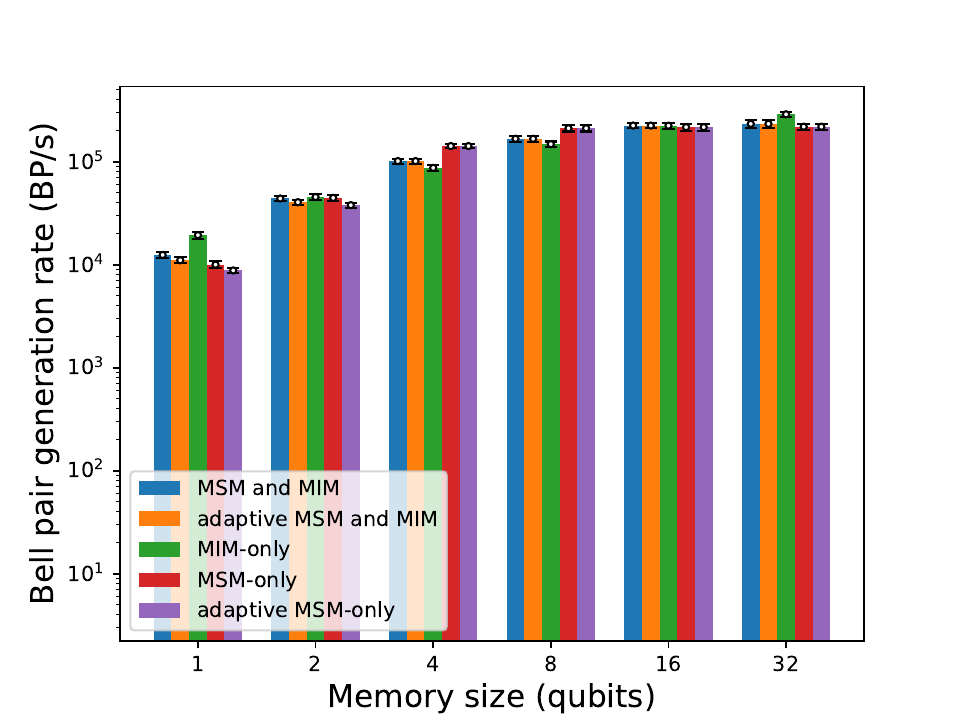}
    \subcaption{1km distance between nodes.}
    \label{fig:1kmswap}
\end{minipage}
\begin{minipage}[h]{0.5\textwidth}
    \centering
    \includegraphics[keepaspectratio, width=\textwidth]{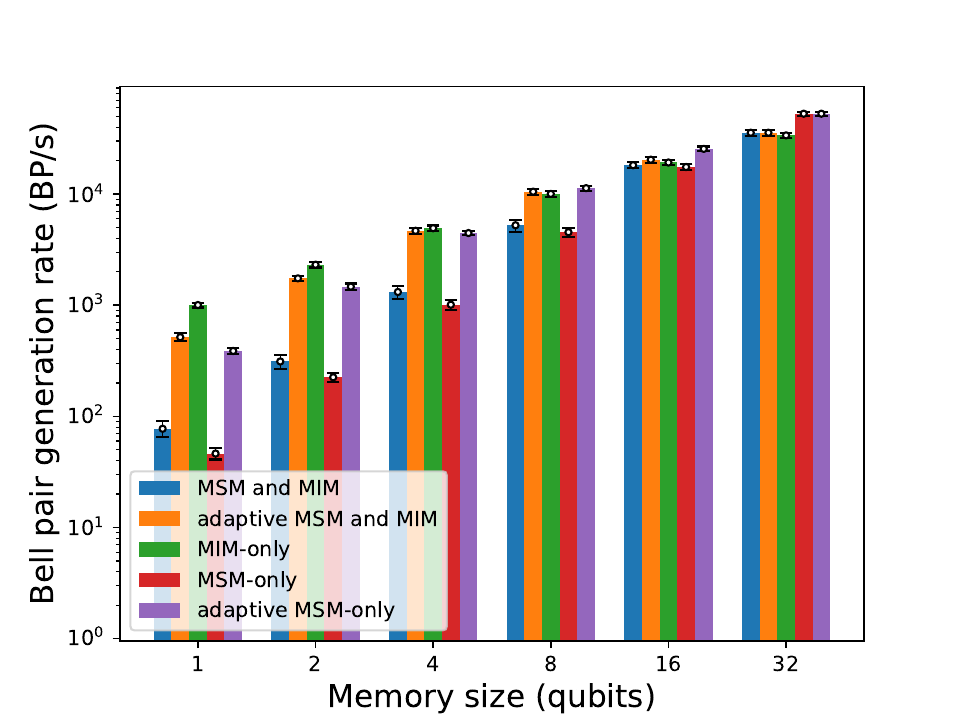}
    \subcaption{20km distance between nodes.}
    \label{fig:20kmswap}
\end{minipage}
\caption{Experiment 2: Number of qubits in quantum memory vs Bell pair generation rate (BP/s) for either 1km or 20km distance between nodes, for two-hop networks, with various mixtures of link architectures. Notice that for either cases the generation rate for "MSM and MIM" resides between MIM-only, and "adaptive MSM and MIM" resides between MIM-only and adaptive MSM-only.}
\label{fig:experiment2}
\end{figure}

Within the MSM links, there are two variations; adaptive and non-adaptive.
We have conducted simulations on these three networks by varying the number of memories in each QNIC and the distance between each node.
The results are shown in Figs.~\ref{fig:1kmswap} (1km) and \ref{fig:20kmswap} (20km).
This result shows that generating Bell pairs through a network under a heterogeneous link architecture is feasible without any large drop in the overall Bell pair generation rate.

We observe that the performance of heterogeneous networks (MIM with MSM) consistently falls between the extremes exhibited by MIM-only and MSM-only configurations (both adaptive and non-adaptive).
While MIM-only performance excels in certain scenarios and falters in others, MSM-only performance follows a similar pattern but in the opposite direction.
This trend holds across most of our simulations, and cases, where this effect is absent, can generally be dismissed as statistically insignificant due to all schemes having similar performance.

We can explain why we observe this as follows.
First, the generation rate for heterogeneous links depends mainly on the slower link.
On one side, where there are fast links, seen from the slow link side, the nodes on the fast link side can freely use entanglement on demand.
We can think of the conditional probability of overall success depending on $p$ where $p$ is the success probability of the slow link.
This means the generation rate can be calculated as ${1}/{\left(t_\text{slow link} + t_\text{ES delay}\right)}$, where $t_\text{slow link}$ is the time to generate a single Bell pair for the slower link, and $t_\text{ES delay}$ is the time lag introduced due to entanglement swapping.
If a slow link is connected with a slow link or a fast link with a fast link, the conditional probability of overall success will not be equal to $p$.
This suggests that the overall generation rate is reasonably slower than the generation rate of a single link. 

If we take notice to the results we have shown in experiment one, we can tell that the interpretation above was correct for most of the cases.
This result is in accordance with the impact of having a weak link in a chain~\cite{van2013path}, which addressed that the performance of a heterogeneous path was essentially limited by the worst link.

\subsection*{\textbf{Experiment 3: Longer Paths}}
\begin{figure*}[h]
    \begin{minipage}[h]{0.5\textwidth}
        \includegraphics[keepaspectratio, width=\textwidth]{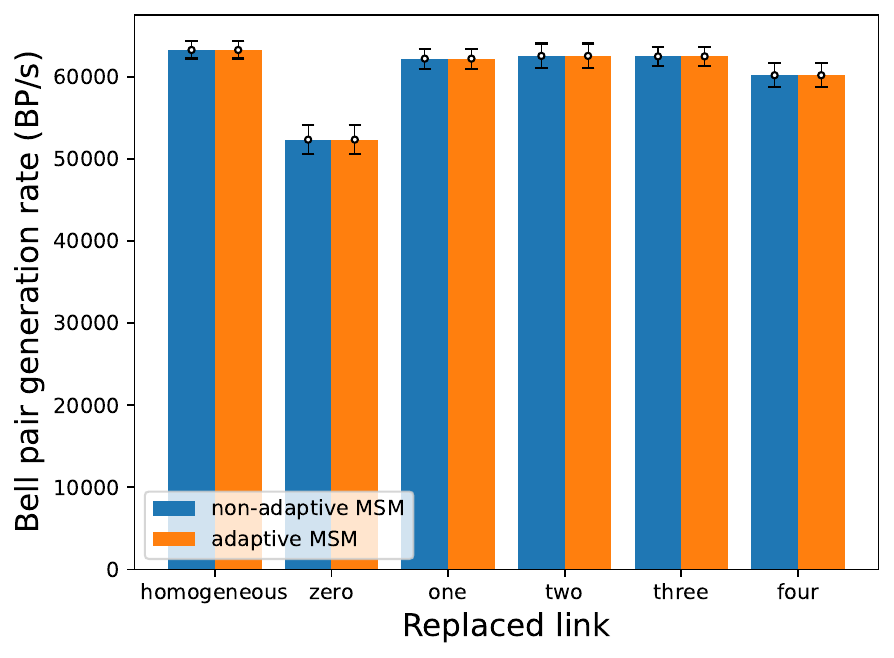}
        \subcaption{1km distance between nodes, for four memory qubits.}
        \label{fig:1kminsertfour}
    \end{minipage}
    \begin{minipage}[h]{0.5\textwidth}
    \includegraphics[keepaspectratio, width=\textwidth]{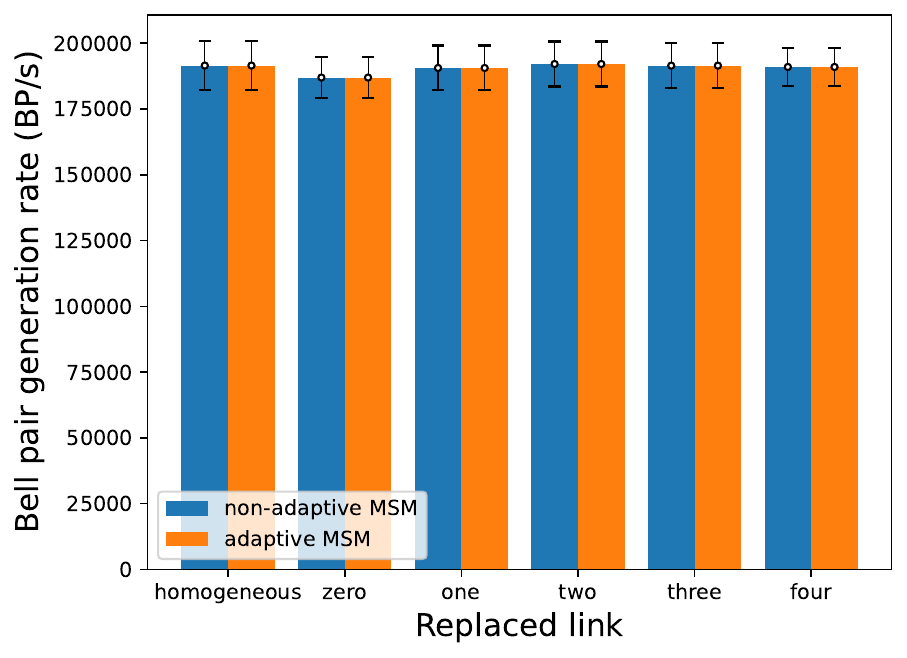}
    \subcaption{1km distance between nodes, for 32 memory qubits.}
    \label{fig:1kminsert32}
\end{minipage}

\begin{minipage}[h]{0.5\textwidth}
    \includegraphics[keepaspectratio, width=\textwidth]{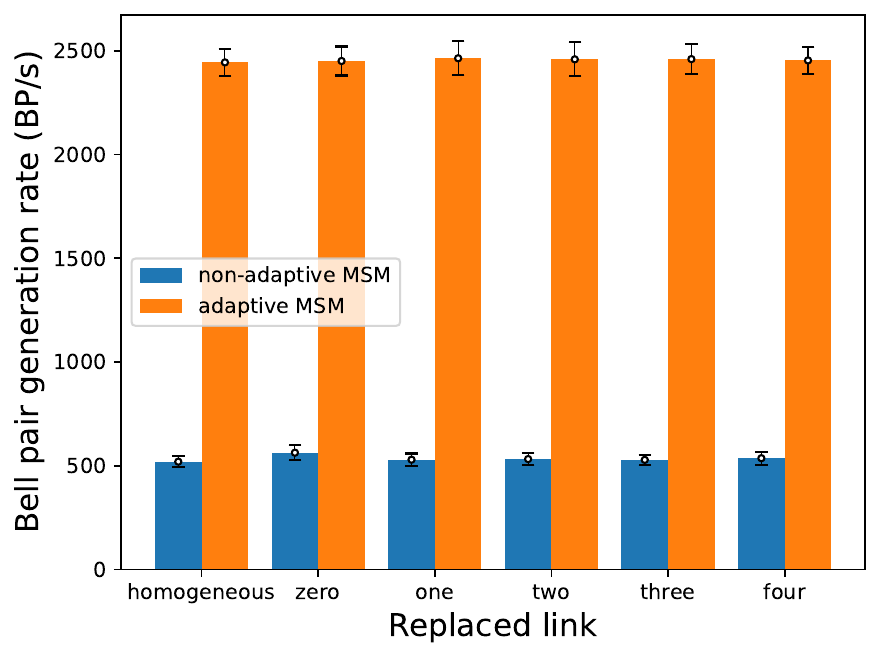}
    \subcaption{20km distance between nodes, for four memory qubits.}
    \label{fig:20kminsertfour}
\end{minipage}
\begin{minipage}[h]{0.5\textwidth}
    \includegraphics[keepaspectratio, width=\textwidth]{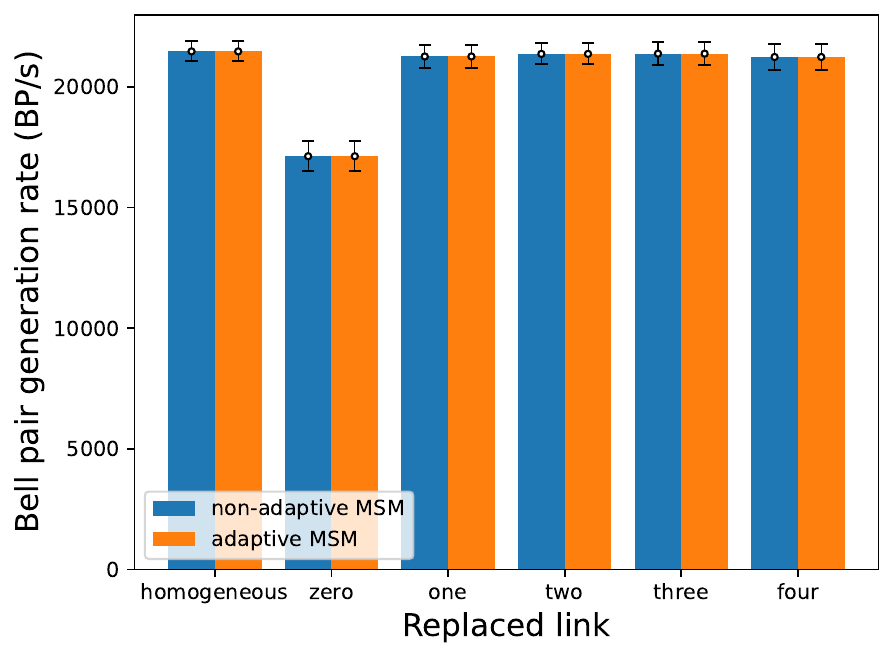}
    \subcaption{20km distance between nodes, for 32 memory qubits.}
    \label{fig:20kminsert32}
\end{minipage}
    \caption{Experiment 3: Location of MIM link in the network vs Bell pair generation rate (BP/s) for ten-hop dominantly MSM networks, where the node distance is either 1km or 20km, and the number of quantum memories is either four or 32.  Notice that for \ref{fig:1kminsertfour} and \ref{fig:20kminsert32}, having link zero replaced, the generation rate slightly decrease, and the homogeneous networks outperform heterogeneous networks in general.}
        \label{fig:experiment3}
\end{figure*}

Next, we would like to investigate whether there will be a difference in entanglement generation when replacing different parts of a homogeneous path with one different link, which makes it a heterogeneous path.
The network is shown in Fig.~\ref{fig:chain_ntwk}, and we replace links zero, one, two, three, or four. In our simulation cases, the majority link is MSM, and the minority link is MIM. We have conducted simulations on different numbers of memories for different link lengths. The x-axis indicates which link from Fig.~\ref{fig:chain_ntwk} is replaced by the minority link type. The results are shown in Figs.~\ref{fig:1kminsertfour} (1km, four memories),~\ref{fig:1kminsert32} (1km, 32 memories),~\ref{fig:20kminsertfour} (20km, 4 memories), and~\ref{fig:20kminsert32} (20km, 32 memories).

Within both homogeneous and heterogeneous cases, most exhibit the same or similar entanglement generation rate.
However, for instances where there is a significant difference in the performance between single MIM links and single MSM links (specifically, for cases where $\mathcal{N}=4$, node distance 1km and $\mathcal{N}=32$, node distance 20km, as also seen in experiment one), replacing link zero leads to a change in the overall entanglement generation rate.
In this case, when MIM is relatively slower than MSM, the entire generation rate also decreases; we see a slight improvement when MIM is fast.
Regarding link four, there is a slight decrease in the generation rate for instances where MIM is slower.

From Fig.~\ref{fig:chain_ntwk}, we can see that some repeater nodes consume their link Bell pairs earlier in the entanglement swapping process, whereas some other nodes, especially the end nodes, need to keep their Bell pair for a fairly long time.
This means that the repeater nodes can generate link-level Bell pairs while waiting for the generation of end-to-end Bell pairs, but the end nodes do not have that buffer time.

This result suggests that improving the entanglement generation rate at the end node and other links where the ``load'' is large, even for homogeneous networks, is likely required for efficient overall Bell pair generation under our entanglement swapping method.
In our case, the critical links were link zero. We observed a similar behavior in link four as well.
We can claim that even if we use methods other than the bisect method to schedule the entanglement swapping, the links with a large load still require an efficient link-level Bell pair.

\section{Conclusion}

We performed several numerical simulations addressing MSM and MIM link heterogeneity, using the quantum internet simulator QuISP.

We have first extended our previous work on the MSM link by observing the effect of adjusting the EPPS pulse rate and compared it to our previous results.
We can notice that the performance of a link strongly depends on the selection of link architecture, and the given number of quantum memories at each node.
We observed that link-level Bell pair generation rate can be seriously degraded for MSM link when the EPPS pulse rate is set at excessively high rate, especially for repeaters with low number of quantum memories.

We introduced an empirical model that describes why this phenomenon occurs, which strongly depends on the EPPS pulse rate.
In the near future, we will have not many quantum memories, where this effect becomes significant. Therefore, adjusting this part was crucial.

Furthermore, we performed several simulations addressing MSM and MIM link heterogeneity.
These quantum link architectures utilize different combinations of optical components.
We demonstrated that we need not do anything special to generate Bell pairs even in the case of heterogeneous networks, and we did not even observe a significant drop in execution time in comparison with MIM-only, MSM-only homogeneous networks.
We introduced a model based on our results, which show that the end-to-end entanglement generation rate strongly depends on the link where it is particularly slow, which depends on the link architecture and the capacity of quantum memories.

Finally, we tested a network consisting of 10 nodes, where the majority link architecture was set to be an MSM link, a minority link as an MIM link, and observed the difference in relocating the MIM link through the network.
We observed that the performance changed when we replaced the end node links and some other critical points in the path.
We described this as a phenomenon dependent on our entanglement swapping order.
If we keep holding the quantum memory, we do not have as much buffer as those who release their memories sooner in the process of entanglement swapping.
Thus, it was natural to observe that replacing the MSM link connected to an end node with an MIM link, especially under conditions where the MIM link was relatively slower than the MSM link, resulted in an overall decrease in performance.
Such data provides insight in optimizing the entanglement swapping sequence, as our simulator is based on the RuleSet architecture.

In analyzing our results, we noticed that the link generation time was the key factor in our observation of such behavior.
This suggests that heterogeneity among memory-based quantum link architectures is still an issue to be addressed, but it does not cause serious generation time loss and is completely feasible.

Our work did not consider multiplexing or heterogeneity in non-memory-based link architectures, and the network configurations were limited to simple cases.
We leave heterogeneity between memory-equipped and memoryless quantum repeaters as future work.

The issue of heterogeneity must be a design goal of a quantum internet architecture from the beginning; acceptable behavior must be designed in from the beginning. This work provides a first look at an expected form of heterogeneity, but we can also expect the unexpected, the introduction of still more previously unknown, heterogeneous technologies over time.  We believe tolerance and robustness to be crucial to deploying a fully scalable quantum internetwork. 

\section*{Code Availability}
The code we used to obtain the results can be found in the GitHub repository for QuISP, under the branch \url{https://github.com/sfc-aqua/quisp/tree/heterogeneous-msmmim}.
\section*{Acknowledgment}
We thank Akihito Soeda, Bernard Ousmane Sane and Amin Taherkhani for valuable discussions.
Grammarly was used to enhance the text quality.
\bibliographystyle{IEEEtran.bst}
\bibliography{biblio}% Produces the bibliography via BibTeX.

\end{document}